\documentclass{PoS}

\usepackage{amsmath}

\title{Phase diagram of scalar field theory on fuzzy sphere and multitrace matrix models}

\ShortTitle{Phase diagram of scalar field theory on fuzzy sphere and multitrace matrix models}

\author{\speaker{J. Tekel}\\
        Department of Theoretical Physics, Faculty of Mathematics, Physics and Informatics\\
				Comenius University\\
Mlynska Dolina 842 48, Bratislava, Slovakia\\
        E-mail: \email{tekel@fmph.uniba.sk}}

\abstract{We study the phase diagram of the scalar field theory on the fuzzy sphere described as a particular multitrace matrix model. We consider perturbative and nonperturbative terms in the kinetic term effective action and describe consequences for the asymmetric regime, the free energy and the location of the triple point within the approximation.}

\FullConference{Proceedings of the Corfu Summer Institute 2015 "School and Workshops on Elementary Particle Physics and Gravity"\\
		1-27 September 2015\\
		Corfu, Greece}

\def\ep{\varepsilon}
\def\half{\frac{1}{2}}

\def\lr#1{\left(#1\right)}

\def\pd#1#2{\frac{\partial #1}{\partial #2}}
\def\slr#1{\left[#1\right]}

\def\trl#1{\textrm{Tr}\lr{#1}}

\def\avg#1{\left\langle #1\right\rangle}

\def\CPn{\mathbb C P^n}

\def\RS2{\mathbb R\times S^2_F}

\def \be  {\begin{equation}}
\def \ee  {\end{equation}}
\def \bex  {\begin{equation*}}
\def \eex  {\end{equation*}}
\def \bea {\begin{eqnarray}}
\def \eea {\end{eqnarray}}
\def \bal {\begin{align}}
\def \eal {\end{align}}

\def\no{\nonumber\\}

\def \PRD {{Phys. Rev. D\ }}
\def \JHEP {{JHEP\ }}

\begin{document}

\section{Introduction}

Th phase structure of noncommtutative field theories has been an active area of research in the past decade. It captures the crucial differences between the commutative and noncommutative theories and allows us to see this difference in an alternative way. Such structure can be studied numerically using Monte Carlo simulations and analytically using matrix model techniques, providing two powerful nonperturbative tools to analyse the properties of noncommutative field theories.

We report on a recent progress in the analytical understanding of the phase structure of the scalar field theory on fuzzy spaces. We deal with theory defined on the fuzzy sphere but most of what we do can be straightforwardly generalized to fuzzy $\mathbb C P^n$. We will show that a nonperturbative approximation to the kinetic term effective action leads to a phase diagram with all the features one expects for the diagram of the fuzzy scalar field theory, most notably the uniform order phase and the triple point.

After a brief overview of the basic notions, previous results and tools to be used we describe two different matrix models describing the scalar $\phi^4$ theory on the fuzzy sphere. We show that the first one, which is a perturbative one, does not lead to a well behaved phase diagram and that the second one does. This report is based on \cite{JT14,JT15} and a thouruough overview of the subject can be found in \cite{mojeakty}.

\section{Fuzzy field theory as a multitrace matrix model}

In this section, we will sumarize the main aspects of the fuzzy field theory and the matrix models techniques used in the rest of the discussion. We will be very brief and we refer the reader to \cite{mojeakty} and references therein for details.

\subsection{Fuzzy field theory}\label{sec2.1}

We will first describe the construction of the scalar field theory on the fuzzy sphere. See \cite{bal} for details and further references.

The fuzzy sphere $S^2_F$ is a space for which the algebra of functions is generated by
\be x_i x_i = \rho^2 \ \ \ , \ \ \ x_i x_j - x_j x_i=i \theta \ep_{ijk} x_k\ .\label{algbr}\ee
These can be realized as a $N=2j+1$ dimensional representation of the $SU(2)$
\be x_i=\frac{2r}{\sqrt{N^2-1}} L_i \ \ \ , \ \ \ \theta=\frac{2r}{\sqrt{N^2-1}} \ \ \ , \ \ \ \rho^2=\frac{4r^2}{N^2-1} j(j+1)=r^2\ .\ee
We see that the coordinates $x_i$ still carry an action of $SU(2)$ and thus the space still has the symmetry of the sphere. We also see that the limit of a large $N$ reproduces the original sphere $S^2$, as the noncommutativity parameter $\theta$ vanishes. Since $x_i$'s are ${N\times N}$ matrices and functions on $S^2_F$ are combinations of their products (elements of the algebra (\ref{algbr})) we come to a conclusion that such functions, and thus scalar fields on $S^2_F$, are given by a herminitian matrices $M$.
	
Derivatives become commutators with generators $L_i$, integrals become traces\footnote{For details about quantization of Poisson manifolds, see \cite{steinpos}.} and we can write an Euclidean field theory action
\begin{align}
	S(M)&=\frac{4\pi R^2}{N}\trl{-\frac{1}{2 R^2} [L_i,M][L_i,M]+\half r M^2+V(M)}=\no&=\trl{\half M[L_i,[L_i,M]]+\half r M^2+V(M)}\ ,\label{action}
\end{align}
where we have absorbed the volume factor into the definition of the field and the couplings. 

The dynamics of the theory is then given by functional correlation functions
\be\left\langle F\right\rangle=\frac{\int dM\, F(M)e^{-S(M)}}{\int dM\,e^{-S(M)}}\ .\label{funccorr}\ee

For our purposes, the most important property stemming from this formula is the trademark property of the noncommutative field theories called the UV/IR mixing \cite{uvir1}. It arises as a consequence of  the non-locality of the theory and renders the commutative limit of noncommutative theory (very) different from the commutative theory we started with. In our setting the UV/IR mixing will result into an extra phase in the phase diagram, not present in the commutative theory and surviving the commutative limit. To conclude this subsection, let us describe the phase structure of both the commutative and noncommutative $\phi^4$.

The commutative field theory has two phases in the phase diagram, disorder and uniform order phases \cite{comr2}. In the first phase, the field oscillates around the value $\phi=0$. In the second phase, the field oscillates around a nonzero value which is the a minimim of the potential, spontaneously breaking the  $Z_2$ symmetry $\phi\to-\phi$ of the theory.
	
\begin{figure}
\centering
\includegraphics[width=0.6\textwidth]{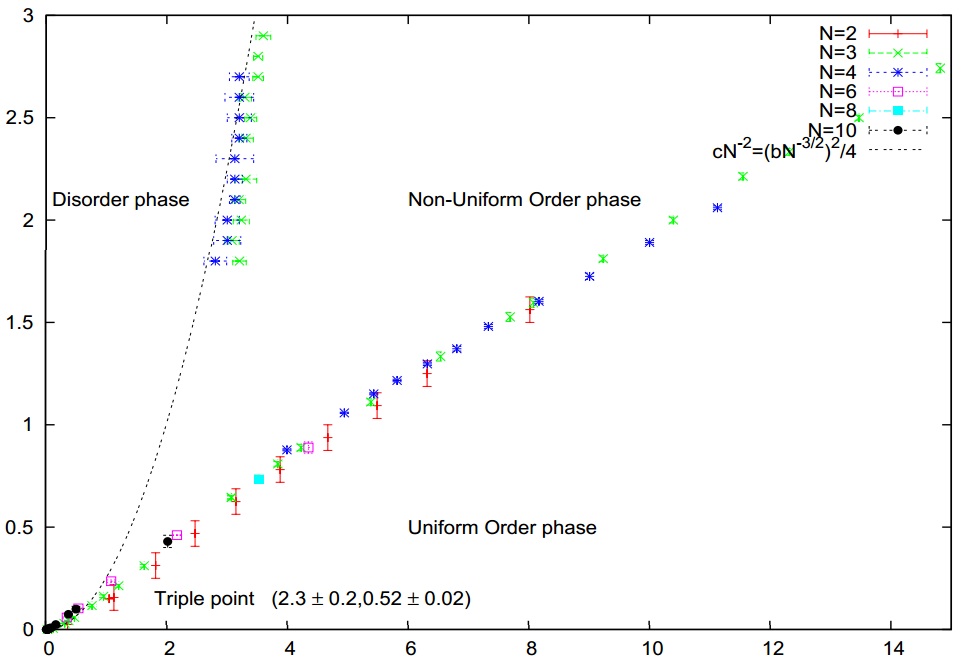}
\caption{The numerical phase diagram of $\phi^4$ theory on the fuzzy sphere from \cite{num09}, presented here with a kind permission of the author.}
\label{fig2.1}
\end{figure}

Noncommutative theories have a third phase, a striped or a nonuniform order phase. In this phase the field does not oscillate around one given value of the field in the whole space, which is a consequence of the nonlocality of the theory and thus is argued to be result of the UV/IR mixing. It also clearly breaks the continuous translational symmetry. Existence of this phase has been established computationally \cite{noncomphase1} and numerically for fuzzy sphere in large body of work \cite{num04,num06,num07,num09,num14}, with the phase diagram of \cite{num09} shown in the figure \ref{fig2.1}.

We see in the figure that the three transition lines meet at a triple point close to the origin of the diagram. The numerical works agree on the critical value of the coupling
\be g_c\approx( 0.125,0.15)\ .\label{numcrit}\ee
Reconstructing this phase diagram and obtaining this value is the main goal of our discussion.

\subsection{Matrix model formulation of fuzzy field theory}

We see that the functional correlation functions of the scalar field theory (\ref{funccorr}) are matrix integrals describing expectations values in a particular matrix model. The probability distribution in this matrix model is determined by the field theory action (\ref{action}). 

Using the standard diagonalization procedure $M=U\Lambda U^{\dagger}$ for some $U\in SU(N)$ and $\Lambda=diag(\lambda_1,\ldots,\lambda_N)$, the integration measure becomes
\be dM=dU \lr{\prod_{i=1}^N d\lambda_i}\times \prod_{i<j}(\lambda_i-\lambda_j)^2\ ,\ee
i.e. the celebrated emergence of the Vandermonde determinant. If we are interested in expectation values of invariant functions, we are to compute integrals like
\be
	\avg{F}\sim\int \lr{\prod_{i=1}^N d\lambda_i}F(\lambda_i)\,e^{-N^2\slr{\half r \sum \lambda_i^2+g \sum \lambda_i^4-\frac{2}{N^2}\sum_{i<j}\log|\lambda_i-\lambda_j|}}\int dU\,e^{-N^2\half \trl{U\Lambda U^{\dagger}[L_i,[L_i,U\Lambda U^{\dagger}]]}}\ .
\ee
We have exponentiated the Vandermonde determinant into the probability distribution and introduced an explicit $N^2$ scaling of the action to obtain a consistent large $N$ limit. This in turn means that the matrix $M$ and the couplings $r,g$ are rescaled by some appropriate power of $N$ to keep everything finite in the large $N$ limit.

The idea is now to treat the angular integral, which is a function of the eigenvalues $\lambda_i$ only, as an effective term in the eigenvalue probability distribution
\begin{align}
	\avg{F}\sim\int \lr{\prod_{i=1}^N d\lambda_i}F(\lambda_i)\,&e^{-N^2\slr{S_{eff}(\lambda_i)+\half r \sum \lambda_i^2+g \sum \lambda_i^4-\frac{2}{N^2}\sum_{i<j}\log|\lambda_i-\lambda_j|}}\ ,\label{mataverage}\\
	&e^{-N^2S_{eff}(\lambda_i)}=\int dU\,e^{-N^2\half \trl{U\Lambda U^{\dagger}[L_i,[L_i,U\Lambda U^{\dagger}]]}}\ .\label{matrixintegral}
\end{align}
We will see in the section \ref{sec3} that this angular integral leads to multitrace terms and thus the matrix models we deal with is a multitrace matrix model.

\subsection{Basic techniques of matrix models}\label{sec2.3}

Finally, let us give some basic tools to deal with integrals line (\ref{mataverage}) and show how one can recover the phase structure of the matrix model from such calculations.

In the large-$N$ limit, which corresponds to the commutative limit, the eigenvalue measure in (\ref{mataverage}) localizes on the extremal configuration $\tilde \lambda_i$ for which
\be \left.\pd{}{\lambda_i}\slr{S_{eff}(\lambda_i)+\half r\sum \lambda_i^2+g \sum \lambda_i^4-\frac{1}{N^2} \sum_{i<j}\log|\lambda_i-\lambda_j|}\right|_{\tilde \lambda}=0\label{saddleeq}\ee
Without the peculiar angular integral, the problem is equivalent to finding an equilibrium of $N$ particles in an external potential $\half r x^2+g x^4$ which logarithmically repel each other due to the Vandermonde term \cite{brezin}. Terms coming from the multitraces in $S_{eff}$ then add further interaction among the particles/eigenvalues.

The saddle point equation (\ref{saddleeq}) is then a condition for the distribution of the eigenvalues of the matrix $M$ in the large $N$ limit of the model considered. It can solved with an assumption on the support of the distribution, giving rise to different solutions depending on the values of the parameters $r,g$.

For a model with no kinetic term the situation is well known and easily understood in the particle analogy. For $r>0$, the potential has a single minimum at $x=0$ and the particles in equilibrium are crowded around this point, spread out due to their repulsion to a finite symmetric interval. We call this a one cut solution. However if we allow $r<0$, the potential has two minima and if $r$ is negative enough, the potential barrier between them is too high and the particles/eigenvalues split into two symmetric intervals. We call this a two cut solution. A third type of solution, an asymmetric one cut solution where all the particles in one of the wells of the potential, is also possible, as long as the well is deep enough to confine the particles. In that case, as in the small temperature limit of the particle analogy, the solution with the lower free energy
\be 
\mathcal{F}=-\frac{1}{N^2}\log Z\ , \ Z=\avg{1}=\slr{S_{eff}(\lambda_i)+\half r\sum \lambda_i^2+g \sum \lambda_i^4-\frac{1}{N^2} \sum_{i<j}\log|\lambda_i-\lambda_j|}_{\lambda\to\tilde \lambda}
\ee
i.e. the more probable solution, is realized in the large $N$ limit. It is easy to see that it is the two cut solution, where the particles are in a position of lower potential and are further apart.

The two transition lines can be computed straightforwardly, the line between the one cut and the two cut solution is given by
\be
r=-4\sqrt{g}
\ee
and the boundary of existence for the asymmetric one cut solution is given by $r=-2\sqrt{15}\sqrt{g}$.

The symmetric one cut phase of the matrix model corresponds to the disorder phase of the field theory it describes. Similarly, the asymmetric one cut phase corresponds to the uniform order phase and the two cut phase to the nonuniform order phase of the field theory.

The multitraces in $S_{eff}$ will deform these lines. Most importantly, if the interaction they introduce is attractive, there is a chance that as long as it is strong enough it can overcome the Vandermonde repulsion, rendering the asymmetric one cut solution stable. Since the model is supposed to describe the field theory, where such a phase exists, we expect this to happen also in the corresponding matrix model.

\section{Perturbative approximation and phase diagram}\label{sec3}

We will first look at a perturbative treatment of the integral (\ref{matrixintegral}) and the phase structure of the corresponding multitrace matrix model. The idea is to expand the exponent in the powers of the kinetic term and then use group theoretic techniques to evaluate the $dU$ integral order by order, eventually reexponentiating the expression. The method has been introduced in \cite{ocon,samann} and the most recent result valid up to the fourth order in the kinetic term is due to \cite{saman2}
	\begin{align}
S_{eff}(M)=&\half\slr{\ep\half \lr{c_2-c_1^2}-\ep^2\frac{1}{24}\lr{c_2-c_1^2}^2+\ep^4\frac{1}{2880}\lr{c_2-c_1^2}^4}-\no&-\ep^4\frac{1}{3456}\Big[\lr{c_4 - 4 c_3 c_1 + 6 c_2 c_1^2 - 3 c_1^4}-2\lr{c_2-c_1^2}^2\Big]^2-\no&-\ep^3\frac{1}{432}\Big[c_3 - 3 c_1 c_2 + 2 c_1^3\Big]^2\ ,\label{pertSeff}\\
c_n=&\trl{M^n}\ ,\nonumber
	\end{align}
where we have introduced an explicit factor $\ep$ multiplying the kinetic term to keep track of the perturbative order. Clearly, this model is beyond any chance of being exactly solvable. We thus solve the saddle point approximation conditions perturbatively, order by order in $\ep$. After some lengthy algebra \cite{JT14,JT15} we obtain the boundary of existence of the symmetric one cut solution
\begin{equation}
r=- 4 \sqrt{g}-\ep\half + \ep^2\frac{1}{12 \sqrt{g}}  +\ep^4 \frac{7}{5760 g^{3/2}}+\ep^6\frac{29}{1935360 g^{5/2}}
\end{equation}
and the boundary of existence for the asymmetric one cut solution
	\begin{align}
	r=&- 2 \sqrt{15} \sqrt{g}+\ep\frac{2}{5} - \ep^2\frac{19}{ 18000 \sqrt{15} \sqrt{g}}
	+\ep^3\frac{29}{1125000 g}\no&- \ep^4\frac{7886183}{4374000000000 \sqrt{15} g^{3/2}}\ .
	\end{align}
Unfortunately, these two lines give a phase diagram which does not reproduce the numerically obtained diagram. They do not intersect giving no triple point and the regions of existence of the different solutions do not match the numerical data even qualitatively.

We thus conclude that perturbative approximations are not good in order to explain the phase structure of the fuzzy scalar field theory and we need to find a way to treat the integral (\ref{matrixintegral}) in a nonperturbative fashion.

\section{Nonperturbative approximation and phase diagram}

We present a nonperturbative treatment of the integral (\ref{matrixintegral}) first used in \cite{poly13} and show its consequences for the phase structure of the theory. This approach is based on an observation that for the free theory with $g=0$, the eigenvalue distribution remains the Wigner semicircle, just with a rescaled radius \cite{steinacker05}.

It can be shown that this means that the kinetic term effective action has to be in the following form \cite{poly13}
\begin{equation}
S_{eff}=\half F(c_2)+\mathcal R=\half \log\lr{\frac{c_2}{1-e^{-c_2}}}+\mathcal R\ ,
\end{equation}
where the remainder $\mathcal{R}$ vanishes for the semicircle distribution. Note, that this agrees with the perturbative expression (\ref{pertSeff}). We also note that if we take the effective action to be $F(c_2-c_1^2)$, we correctly recover also the odd terms. So as an approximation, we neglect the remainder term and consider the model with $S_{eff}=\half F\lr{c_2-c_1^2}$.

The symmetric regime $c_1=0$ of this model is not too difficult to solve and the phase transittion line between the symmetric one and two cut regimes is given by \cite{poly13}
\be r=-5\sqrt{g}-\frac{1}{1-e^{1/\sqrt g}}\ .\label{4symtr}\ee
Note that at the phase transition is at $c_2=1/\sqrt g$, which explains the problems of perturbative expansion, which is effectively a small moment expansion. The expressions we encounter are not analytic and perturbation theory is bound to fail.

The asymmetric regime of the model can not be solved analytically, simply because the saddle point approximation leads to polynomial equations of a high degree \cite{mojeakty}. To proceed in a nonperturbative fashion, we solve the equations numerically \cite{JT15}. 

Once the equations are solved, we find out that in a region of the parameter space no asymmetric solution is possible. In the rest of the parameter space, different types of asymmetric one and two cut solutions are possible. After recalling the free energy criterion from the section \ref{sec2.3}, we find out that the completely asymmetric one cut solution is the preferred solution wherever an asymmetric solution exists.

\begin{figure}
\centering
\includegraphics[width=0.48\textwidth]{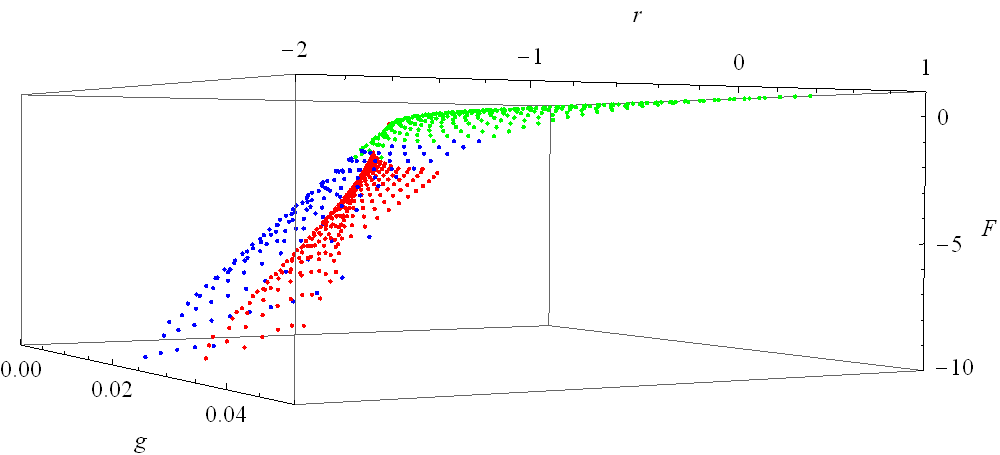}
\includegraphics[width=0.48\textwidth]{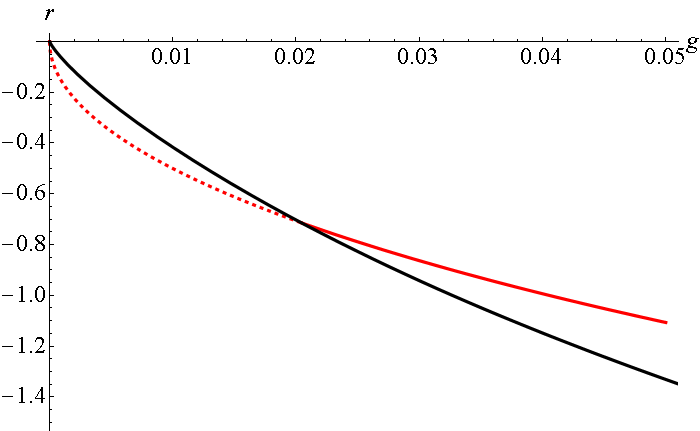}
\caption{The left figure shows the comparison of free energies of symmetric one cut (green), asymmetric onecut (red) and symmetric two cut (blue) solutions. The right figure showsThe phase diagram of the nonperturbative approximation. The red line is the line (\protect\ref{4symtr}), the black line is the numerically obtained boundary of existence of the asymmetric one cut solution.}
\label{fig4.1}
\end{figure}

Moreover, when we compare its free energy to the free energy of the symmetric one cut and two cut solutions we find out that it is the preferred solution. After numerically identifying the edge of the region of exstence of the asymmetric solution, we obtain the phase diagram of the theory shown in the figure \ref{fig4.1}.

From the figure, we see that the diagram has the desired qualitative features of the numerical phase diagram mentioned in the section \ref{sec2.1}. The location of the triple point is then given by the critical coupling
\be g_c\approx 0.02\ ,\ee
which differs from the numerical value (\ref{numcrit}) by a factor of roughly 7. Let us conclude this section by commenting on this discrepancy, arguing that there is room for improvement on both the numerical and analytical side.

In our approximation, we have neglected all the higher multitrace terms in the remainder $\mathcal{R}$. Their addition will deform the transition lines in the figure \ref{fig4.1} and shift the location of the triple point. The numerical results on the other hand use a linear extrapolation of data quite far from the predicted location of the triple point. Since the transition lines curve considerably close to the triple point, it is natural to expect the linear extrapolation to lead to some inaccuracy. After some preliminary analysis of numerical data, it seems to be the case, but it is not clear to what extent \cite{denjoepersonal}.

\section{Conclusions and outlook}

The main conclusion of this report is twofold. First, the triple point of the $\phi^4$ theory on the fuzzy sphere is in the region of parameter space where the kinetic term must be treated nonperturbatively and perturbative expansion leads to inconsistent results. Second, even though the kinetic term can not yet be treated compeltely, a particular nonperturbative approximation does lead to model with consistent phase structure compatible with the previous numerical results.

These results leave several interesting and important goals for future research. A more complete and perhaps the complete treatment of the angular integral (\ref{matrixintegral}) will lead to a better description of the phase structure of the theory. Generalization of these ideas to the noncommutative plane \cite{num14panero2} and other noncommutative spaces where numerical data is available, for example the fuzzy torus \cite{fuzzydiscnum} or $\mathbb R\times S_F^2$ \cite{rsfnum}, can further cross check this method with the nummerical approach. And the most important goal is the understanding the phase strucuture of a theory free of the UV/IR mixing \cite{uvir}. It has been argued that the striped phase is a consequence of the mixing a thus this phase is expected to be missing from the phase diagram.

\acknowledgments{
I would like to thank Denjoe O'Connor for useful discussions and organizers of the \emph{Corfu Summer Institute 2015} for a wonderful conference and for a opportunity to present the results of my research.

My stay at the \emph{Corfu Summer Institute 2015} has been supported by the COST Action MP1405 QSPACE, supported by COST (European Cooperation in Science and Technology). This work was supported by the \emph{Alumni FMFI} foundation as a part of the \emph{N\'{a}vrat teoretikov} project.
}

\end{document}